\documentclass[prl,aps,showpacs,twocolumn,floatfix]{revtex4}
\usepackage{epsfig} \usepackage{graphics} \usepackage{bm}
\usepackage{amssymb}
\usepackage{graphicx}
\begin{document}

\preprint{Lebed-PRL}

\title{A Chiral Triplet Quasi-Two-Dimensional Superconductor in a Parallel
Magnetic Field}

\author{A.G. Lebed$^*$}

\affiliation{Department of Physics, University of Arizona, 1118 E.
4-th Street, Tucson, AZ 85721, USA}

\begin{abstract}
We calculate the parallel upper critical magnetic field
$H_{\parallel}(0)$ for an in-plane isotropic quasi-two-dimensional
(Q2D) chiral triplet superconductor at zero temperature, $T=0$. In
particular, the ratio
$H_{\parallel}(0)/(|dH^{GL}_{\parallel}/dT|_{T=T_c}T_c) = 0.815$
is defined, where $|dH^{GL}_{\parallel}/dT|_{T=T_c}$ is the
so-called Ginzburg-Landau slope of the upper critical magnetic
field, $T_c$ is a superconducting transition temperature at $H=0$.
We show that the theoretically obtained above mentioned value
strongly contradicts to the experimentally measured ones in a
candidate for a chiral triplet superconductivity Sr$_2$RuO$_4$,
which provides one more argument against the chiral triplet
scenario of superconductivity in this compound. Our results may be
useful for establishing chiral triplet superconductivity in other
Q2D candidates for this phenomenon.
\end{abstract}


\maketitle

Since discovery of superconductivity in the quasi-two-dimensional
(Q2D) conductor Sr$_2$RuO$_4$ [1], it has been intensively
investigated for more than 25 years (for reviews, see Refs.[2,3]).
Some analogy of this Q2D superconductor with the superfluid $^3$He
was recognized from the beginning and the existence of a chiral
triplet superconducting phase in the Sr$_2$RuO$_4$ was suggested
[4]. This scenario of superconducting pairing was supported by the
observations of no change of the Knight shift between normal and
superconducting phases [5,6] and breaking of the time reversal
symmetry in the superconducting phase [7,8]. On the other hand,
there were arguments against the chiral triplet superconductivity
scenario, which were almost ignored that time by scientific
community. One of the first argument was the paramagnetic
limitation of the parallel upper critical magnetic field in
Sr$_2$RuO$_4$ [9,10]. In addition, the predicted in the chiral
triplet scenario edge currents were not found in the Sr$_2$RuO$_4$
[11,12] but were found zeros of superconducting gap on Q2D Fermi
surface (FS) [13,14], which is against the fully gaped chiral
triplet scenario [4]. Recently, the strongest experimental
argument against the triplet scenario of superconductivity in
Sr$_2$RuO$_4$ was published [15], where strong drop of the Knight
shift in superconducting state of the above mentioned material was
experimentally discovered.

As seen from the above discussion, the situation with the chiral
triplet scenario of superconductivity in Sr$_2$RuO$_4$ is still
rather controversial. The goal of our Letter is two-fold. First,
we improve and make our pioneering argument [9] in favor of
singlet superconductivity in Sr$_2$RuO$_4$ to be firm. The point
is that in Ref.[9] (see also recent Ref.[16]) we calculated the
ratio $H_{\parallel}(0)/(|dH^{GL}_{\parallel}/dT|_{T=T_c}T_c) =
0.75$, where $|dH^{GL}_{\parallel}/dT|_{T=T_c}$ is the so-called
Ginzburg-Landau (GL) slope, for $s$-wave Q2D superconductivity and
compared it with the experimental one, 0.45-0.5 [17-19]. To make
our argument against the chiral triplet scenario to be firm, below
we calculate the above mentioned ratio exactly for the in-plane
isotropic chiral triplet superconductor with ${\bf d}$ vector
order parameter [4,20],
\begin{equation}
{\bf d}= {\bf z} \ \Delta_0 \ (k_x \pm i k_y),
\end{equation}
and obtain even stronger inconsistency,
\begin{equation}
H_{\parallel}(0) = 0.815 \ |dH^{GL}_{\parallel}/dT|_{T=T_c} \ T_c,
\end{equation}
with the experimental values [17-19], where, to the best of our
knowledge, Eq.(2) is the first time obtained in the Letter. The
second our goal is to suggest one more test for chiral triplet
superconductivity, which may already exist in the slightly
in-plane anisotropic Q2D triplet superconductor UTe$_2$ [21] and,
as we hope, will be discovered in some other Q2D compounds in the
future.

Note that superconducting phase (1) is not destructed by the Pauli
paramagnetic spin-splitting effects in a parallel magnetic field
[9,20,22]. Moreover, in accordance with Refs.[23-28], it has also
to restore superconductivity in very high magnetic fields due to
quantum effects of the Bragg reflections from boundaries of the
Brillouin zone. In this Letter, we consider the so-called
quasi-classical parallel orbital upper critical magnetic field
[29,16,30], $H_{\parallel}(0)$, which destroys superconductivity
in Q2D superconductors at low temperatures in an intermediate
region of the fields.

Let us consider a layered superconductor with the following
in-plane isotropic Q2D electron spectrum:
\begin{equation}
\epsilon({\bf p})= \epsilon (p_x,p_y) - 2 t_{\perp} \cos(p_z c^*),
\ \ \  t_{\perp} \ll \epsilon_F,
\end{equation}
where
\begin{equation}
\epsilon (p_x,p_y) = \frac{ (p^2_x + p^2_y)}{2m} \ , \ \ \
\epsilon_F = \frac{p^2_F}{2m} \ .
\end{equation}
[In Eqs.(3) and (4), $t_{\perp}$ is the integral of overlapping of
electron wave functions in a perpendicular to the conducting
planes direction, $m$ is the in-plane electron mass, $\epsilon_F$
and $p_F$ are the Fermi energy and Fermi momentum, respectively;
$\hbar \equiv 1$.] In a parallel magnetic field, which is applied
along ${\bf x}$ axis,
\begin{equation}
{\bf H} = (H,0,0) \ ,
\end{equation}
it is convenient to choose the vector potential of the field in
the form:
\begin{equation}
{\bf A} = (0,0,Hy) \ .
\end{equation}
For electron motion within the conducting planes and between the
planes, we make use of the the so-called Peierls substitution
method [22] in the electron energy spectrum (3) and (4):
\begin{equation}
p_x \rightarrow - i \biggl( \frac{\partial }{ \partial x} \biggl),
\ \ p_y \rightarrow - i \biggl( \frac{\partial }{
\partial y} \biggl),
\end{equation}
and
\begin{equation}
p_z c^* \rightarrow p_z c^*- \biggl( \frac{\omega_c}{v_F} \biggl)
y, \ \ \omega_c = \frac{e v_F c^* H}{c},
\end{equation}
where $e$ is the electron charge, $c$ is the velocity of light.

After the Peierls substitutions in a magnetic field (7) and (8),
the electron Hamiltonian (3) and (4) becomes:
\begin{equation}
\hat H = - \frac{1}{2m} \biggl( \frac{\partial^2 }{ \partial x^2}
+ \frac{\partial^2 }{ \partial y^2} \biggl) - 2 t_{\perp} \cos
\biggl(p_z c^* - \frac{\omega_c}{v_F}y \biggl).
\end{equation}
If we take into account that $\omega_c \ll \epsilon_F$, we can
define the following electron wave functions in the mixed, $(p_x,
y, p_z)$, representation:
\begin{eqnarray}
&\Psi^{\pm}_{\epsilon}(x,y,z) =  \ \exp( i p_x x) \ \exp[\pm i
p^0_y(p_x) y] \ \exp( i p_z z)
\nonumber\\
&\times \psi_{\epsilon}^{\pm}(p_x,y,p_z), \ \  p^0_y(p_x) =
\sqrt{p_F^2-p^2_x},
\end{eqnarray}
where energy $\epsilon$ is counted from the Fermi energy
$\epsilon_F$. Note that Eq.(9) is very general. For instance, it
takes into account the quantum effects for open electron orbits in
a magnetic field - the Bragg reflections from boundaries of the
Brillouin zones (see, Refs. [23-28]). As we already mentioned,
below we calculate the so-called quasi-classical upper critical
magnetic field [29,16,30] and, thus, do not take into account
possible stabilization of superconductors in very high magnetic
fields [23-28], which is possibly observed in the Q2D
superconductor UTe$_2$.

It is easy to prove that, in this case, we can represent the
electron Hamiltonian (9) for the wave functions
$\psi_{\epsilon}^{\pm}(p_x,y,p_z)$ in Eq.(10) as
\begin{eqnarray}
&\biggl\{ \frac{1}{2m} \biggl[p^2_F \pm 2 i p^0_y(p_x)
\frac{d}{dy}\biggl] - 2 t_{\perp} \cos \biggl(p_z c^* -
\frac{\omega_c}{v_F} y \biggl) \biggl\}
\nonumber\\
&\times \psi_{\epsilon}^{\pm}(p_x,y,p_z) = (\epsilon +
\epsilon_F)\ \psi_{\epsilon}^{\pm}(p_x,y,p_z).
\end{eqnarray}
Note that it is easy to represent Eq.(11) in the following more
convenient way:
\begin{eqnarray}
&\biggl[ \pm i v^0_y(p_x) \frac{d}{dy} - 2 t_{\perp} \cos
\biggl(p_z c^* - \frac{\omega_c}{v_F} y \biggl) \biggl]
\psi_{\epsilon}^{\pm}(p_x,y,p_z)
\nonumber\\
&= \epsilon \ \psi_{\epsilon}^{\pm}(p_x,y,p_z), \ \ \ v^0_y(p_x) =
p^0_y(p_x)/m .
\end{eqnarray}
We point out that Eq.(12) is still too general. For instance, for
a pure case, it contains the above discussed quantum effects of an
electron motion in a magnetic field. As mentioned before, below,
we study the quasi-classical case, where, as shown in Refs. [16]
and [30], it is possible to take into account only linear term
with respect to magnetic field. As a result, we obtain:
\begin{eqnarray}
&\biggl[ \pm i v^0_y(p_x) \frac{d}{dy}  -2t_{\perp} \cos (p_zc^*)
-\biggl( \frac{2t_{\perp} \omega_c y}{v_F} \biggl) \sin (p_zc^*)
\biggl]
\nonumber\\
&\times \ \psi_{\epsilon}^{\pm}(p_x,y,p_z) = \epsilon \
\psi_{\epsilon}^{\pm}(p_x,y,p_z).
\end{eqnarray}

In this Letter, we use the so-called Matsubara's variant of the
Green's functions, $G^{\pm} (i\omega_n; p_x; y, y_1 ;p_z)$, method
[31] for interacting electrons in the mixed $(p_x,y,p_z)$
representation. For non interacting electrons in the magnetic
field, equation for Matsubara's Green's function, can be written
as [32]
\begin{eqnarray}
&\biggl[i \omega_n  \mp i v^0_y(p_x) \frac{d}{dy}  +2t_{\perp}
\cos (p_zc^*) +\biggl( \frac{2t_{\perp} \omega_c y}{v_F} \biggl)
\sin (p_zc^*) \biggl]
\nonumber\\
&\times G^{\pm}(i \omega_n;p_x; y,y_1;p_z) = \delta(y-y_1) \ ,
\end{eqnarray}
where $\omega_n$ is the so-called Matsubara's frequency [31]. It
important that Eq.(14) can be solved analytically [16,25]. As a
result, we obtain:
\begin{eqnarray}
&G^{\pm} (i \omega_n; p_x; y, y_1; p_z) = - i \frac{ sgn(
\omega_n)}{v^0_y(p_x)}
 \exp \biggl[ \pm \frac{\omega_n (y-y_1)}{v^0_y(p_x)} \biggl]
\nonumber\\
&\times \exp \biggl[\mp i \frac{2 t_{\perp}\cos (p_zc^*)
(y-y_1)}{v^0_y(p_x)} \biggl]
\nonumber\\
&\times \exp \biggl[ \mp i \frac{t_{\perp} \omega_c
(y^2-y^2_1)}{v^0_y(p_x) v_F} \sin(p_z c^*) \biggl] .
\end{eqnarray}

Let us consider the so-called Gor'kov's gap equation for two
component order parameter (see, for example, Ref. [33]),
\begin{equation}
\Delta({\bf k},{\bf q}) = \Delta_x({\bf q}) \delta_x({\bf k})+
\Delta_y({\bf q}) \delta_y({\bf k}),
\end{equation}
where $[\phi_x({\bf k}),\phi_y({\bf k})]$ are basis functions of
two-dimensional representation of tetragonal group transforming as
$(k_x,k_y)$. The Gor'kov's equation in this case can be written as
[33]:
\begin{eqnarray}
\Delta({\bf k},{\bf q})= T \sum_n \int\frac{d^3 {\bf
k_1}}{(2\pi)^3} \ V \sum_{i=x,y}\delta_i({\bf k})\delta_i({\bf
k_1})
\nonumber\\
\times G^{\pm}(i\omega_n; {\bf k_1}) \ G^{\mp}(-i\omega_n;{\bf
-k_1+q}) \Delta({\bf k_1},{\bf q}).
\end{eqnarray}
Using Fourier transformations of the known Green's functions (15),
it is possible to derive the so-called gap equation in the mixed
representation, determining the upper critical magnetic field for
two component order parameter (16). Below, we use the Q2D Fermi
surface (3) with isotropic in-plane electron 2D spectrum (4) and
the following isotropic chiral triplet electron electron
superconducting interactions:
\begin{eqnarray}
&V \sum_{i=x,y}\delta_i({\bf k})\delta_i({\bf k_1})
\nonumber\\
&= 2g [\cos \phi \cos \phi_1 + \sin \phi \sin \phi_1]= 2 g \cos
(\phi - \phi_1).
\end{eqnarray}
As a result of lengthly but straightforward calculations, we
obtain the Gor'kov's gap equation in the form:
\begin{eqnarray}
&\Delta(\phi, y) = \int_0^{2 \pi} \frac{d \phi_1}{2 \pi} g
\cos(\phi - \phi_1) \int^{\infty}_{ |y-y_1| > d |\sin \phi_1|}
\nonumber\\
&\times \frac{2 \pi T dy_1}{v_F |\sin \phi_1| \sinh \biggl( \frac{
2 \pi T |y-y_1|}{ v_F |\sin \phi_1|} \biggl) }
\nonumber\\
&\times J_0 \biggl[ \frac{2 t_{\perp} \omega_c}{v^2_F |\sin
\phi_1|} (y^2-y^2_1)\biggl]   \ \Delta(\phi_1, y_1)\ ,
\end{eqnarray}
where  $g$ is the effective electron coupling constant, $d$ is the
cut-off distance, $J_0(...)$ is the zero order Bessel function. In
Eq.(19) the superconducting gap $\Delta(\phi,y)$ depends on a
center of mass of the BCS pair, $y$, and on the position on the
cylindrical FS (4), where $\phi$ and $\phi_1$ are the polar angles
counted from ${\bf x}$ axis.

Using the general Eq.(16), we need to seek two solutions of
Eq.(19) in the form,
\begin{equation}
\Delta(\phi,y) = \Delta_x(y) \cos(\phi) + \Delta_y(y) \sin(\phi),
\end{equation}
for given direction of a magnetic field. Since we consider the
isotropic 2D FS (4) and the isotropic chiral triplet
electron-electron interactions (18), it is possible to make sure
that the in-plane upper critical magnetic field does not depend on
direction of the field. Therefore, we can consider magnetic field
applied along ${\bf x}$ axis (5), as we suggested before. For such
magnetic field, it is easy to show that there are the following
two solutions:
\begin{equation}
\Delta_1(\phi,y) = \Delta_x(y) \cos(\phi)
\end{equation}
and
\begin{equation}
\Delta_2(\phi,y) = \Delta_y(y) \sin(\phi),
\end{equation}
which obey the following integral equations:
\begin{eqnarray}
&\Delta_x(y) = \int_0^{2 \pi} \frac{d \phi_1}{2 \pi} g
\cos^2(\phi_1) \int^{\infty}_{ |y-y_1| > d |\sin \phi_1|}
\nonumber\\
&\times \frac{2 \pi T dy_1}{v_F |\sin \phi_1| \sinh \biggl( \frac{
2 \pi T |y-y_1|}{ v_F |\sin \phi_1|} \biggl) }
\nonumber\\
&\times J_0 \biggl[ \frac{2 t_{\perp} \omega_c}{v^2_F |\sin
\phi_1|} (y^2-y^2_1)\biggl]   \ \Delta_x(y_1)\ ,
\end{eqnarray}
and
\begin{eqnarray}
&\Delta_y(y) = \int_0^{2 \pi} \frac{d \phi_1}{2 \pi} g
\sin^2(\phi_1) \int^{\infty}_{ |y-y_1| > d |\sin \phi_1|}
\nonumber\\
&\times \frac{2 \pi T dy_1}{v_F |\sin \phi_1| \sinh \biggl( \frac{
2 \pi T |y-y_1|}{ v_F |\sin \phi_1|} \biggl) }
\nonumber\\
&\times J_0 \biggl[ \frac{2 t_{\perp} \omega_c}{v^2_F |\sin
\phi_1|} (y^2-y^2_1)\biggl]   \ \Delta_y( y_1)\ ,
\end{eqnarray}
correspondingly. Note that the solutions (21) and (22) degenerate
in the absence of a magnetic field. In the presence of the
magnetic field (5) with the vector potential (6), Eqs.(23) and
(24) are not degenerated anymore and correspond to different
solutions, $\Delta_x(y) \neq \Delta_y(y)$, with different upper
critical magnetic fields, $H^x_{\parallel} \neq H^y_{\parallel}$.
The actual upper critical field will be the maximum value of them.
Therefore, below, we need to solve and study the both integral
equations, (23) and (24).

Here, we introduce more convenient variable of the integrations,
\begin{equation}
y_1 -y = z|\sin \phi_1|,
\end{equation}
and rewrite the integral equations (23) and (24) in the following
ways:
\begin{eqnarray}
&\Delta_x(y) = \int_0^{2 \pi} \frac{d \phi_1}{2 \pi} g
\cos^2(\phi_1) \int^{\infty}_{ |z| > d}  \frac{2 \pi T dz}{v_F
\sinh \biggl( \frac{ 2 \pi T |z|}{ v_F} \biggl) }
\nonumber\\
&\times J_0 \biggl[ \frac{2 t_{\perp} \omega_c}{v^2_F}
[z(2y+z|\sin \phi_1|)]\biggl]
\nonumber\\
&\times\Delta_x (y+z|\sin \phi_1|)
\end{eqnarray}
and
\begin{eqnarray}
&\Delta_y(y) = \int_0^{2 \pi} \frac{d \phi_1}{2 \pi} g
\sin^2(\phi_1) \int^{\infty}_{ |z| > d}  \frac{2 \pi T dz}{v_F
\sinh \biggl( \frac{ 2 \pi T |z|}{ v_F} \biggl) }
\nonumber\\
&\times J_0 \biggl[ \frac{2 t_{\perp} \omega_c}{v^2_F}
[z(2y+z|\sin \phi_1|)]\biggl]
\nonumber\\
&\times\Delta_y(y+z|\sin \phi_1|).
\end{eqnarray}

Let us first study Eq.(26). Here, we derive the GL equation for
(26) to determine the so-called GL slope of the upper critical
magnetic field, $|d H^{x(GL)}_{\parallel}/dT|_{T=T_c}$. In order
to derive the GL equation, we expand the superconducting gap and
the Bessel function  in Eq.(26) with respect to small parameter,
$|z| \ll v_F/(\pi T_c)$:
\begin{eqnarray}
&&\Delta_{x}(y+z |\sin \phi_1|) \approx \Delta_{x}(y) +
\frac{1}{2} z^2 \sin^2 \phi_1 \biggl[ \frac{d^2
\Delta_{x}(y)}{dy^2} \biggl],
\nonumber\\
 &&J_0 \biggl\{ \frac{2 t_{\perp} \omega_c}{v^2_F} [z (z |\sin
 \phi_1|
+2y)] \biggl\} \approx 1-\frac{4 t^2_{\perp} \omega^2_c}{v^4_F}
z^2 y^2 .
\end{eqnarray}
Next step is to substitute the obtained expansions (28) into the
integral gap equation (26) and average over the polar angle
$\phi_1$:
\begin{eqnarray}
&&-\frac{1}{8} \biggl[ \frac{d^2 \Delta_{x}(y)}{dy^2} \biggl]
\int_0^{\infty} \frac{2 \pi T_c z^2 d z}{v_F \sinh \biggl( \frac{
2 \pi T_c z}{ v_F} \biggl)}
\nonumber\\
&&+ y^2 \Delta_{x}(y) \frac{4t^2_{\perp}\omega^2_c}{v^4_F}
\int_0^{\infty} \frac{2 \pi T_c z^2 d z}{v_F \sinh \biggl( \frac{
2 \pi T_c z}{ v_F} \biggl)}
\nonumber\\
&&+\Delta_{x}(y) \biggl[ \frac{1}{g} - \int_d^{\infty} \frac{2 \pi
T d z}{v_F \sinh \biggl( \frac{ 2 \pi T z}{ v_F} \biggl)}
\biggl]=0.
\end{eqnarray}
Note that in the absence of a magnetic field the superconducting
transition temperature, $T_c$, obeys the following equation:
\begin{equation}
 \frac{1}{g} = \int_d^{\infty}  \frac{2 \pi T_c d z}{v_F
\sinh \biggl( \frac{ 2 \pi T_c z}{ v_F} \biggl)} \ .
\end{equation}
After substitution of Eq.(30) into Eq.(29), we obtain the
following differential equation:
\begin{equation}
- \xi^2_{\parallel} \biggl[ \frac{d^2 \Delta_{x}(y)}{dy^2} \biggl]
+ \biggl(\frac{2\pi H}{\phi_0}\biggl)^2 \xi^2_{\perp} y^2
\Delta_{x}(y) -\tau \Delta_{x}(y) = 0,
\end{equation}
which is called the GL one. During the derivation of the GL
equation, determining the upper critical magnetic field slope near
$T_c$, $\tau = (T_c-T)/T_c \ll 1$, we introduce the so-called
parallel and perpendicular GL coherent lengths:
\begin{equation}
\xi_{\parallel} = \frac{\sqrt{7 \zeta(3)}v_F}{8 \pi T_c}, \ \ \
\xi_{\perp} = \frac{\sqrt{7 \zeta(3)} t_{\perp} c^*}{2 \sqrt{2}
\pi T_c}.
\end{equation}
Note that $\zeta(x)$ is the so-called Riemann function [34],
\begin{equation}
\int^{\infty}_0 \frac{z^2 dz}{\sinh(z)}  = \frac{7}{3} \zeta(3),
\end{equation}
whereas $\phi_0 = \frac{\pi c}{e}$ is the magnetic flux quantum.
It is important that $\xi_{\parallel}$ in Eqs.(31) and (32) is
different from that in Refs.[16,30], since in this Letter we
consider two-component superconducting gap. To obtain the parallel
upper critical field for Eq.(26) near T$_c$, it is necessary to
solve the Scr\"{o}dinger-like equation (31) and to find its lowest
energy level. Since mathematically Eq.(31) is equivalent to the
Scr\"{o}dinger equation for a harmonic oscillator, we obtain the
following equation for the parallel upper critical field in the GL
region, $\tau \ll 1$:
\begin{equation}
H^{x(GL)}_{\parallel}(T) = \biggl( \frac{\phi_0}{2 \pi
\xi_{\parallel} \xi_{\perp}} \biggl)\tau = \biggl[ \frac{8
\sqrt{2} \pi^2 c T^2_c}{7 \zeta(3) e v_F t_{\perp} c^*} \biggl]
\tau ,
\end{equation}
which, as expected, differs from that in Refs.[16,30].

More complicated problem is to solve Eq.(26) at $T=0$ and, thus,
to find the upper critical magnetic field at zero temperature,
$H^x_{\parallel}(0)$. This is possible to do only by means of
numerical calculations. To this end, let us introduce more
convenient variables,
\begin{equation}
\tilde z = \frac{\sqrt{2t_{\perp} \omega_c}}{v_F} z , \ \ \ \tilde
y = \frac{\sqrt{2t_{\perp} \omega_c}}{v_F} y ,
\end{equation}
and rewrite the gap Eq.(26) for the case $T=0$:
\begin{eqnarray}
&\tilde \Delta_{x}(\tilde y) = g \int_{-\pi/2}^{\pi/2}
\frac{d\phi_1}{\pi} \int^{\infty}_d \frac{d \tilde z}{\tilde z}
J_0 [\tilde z(2 \tilde y + \tilde z \sin \phi_1)]
\nonumber\\
&\times 2 \cos^2(\phi_1) \tilde \Delta_{x}(\tilde y + \tilde z
\sin \phi_1) \ ,
\end{eqnarray}
where
\begin{equation}
\tilde \Delta_{x} (\tilde y) = \Delta \biggl(\frac{v_F \tilde
y}{\sqrt{2 t_{\perp} \omega_c}} \biggl).
\end{equation}

Here, we summarize procedure of the numerical solution of the
integral Eq.(36) and obtain the following new result:
\begin{equation}
H^x_{\parallel}(0) = 10.78 \frac{c T^2_c}{e v_F t_{\perp} c^*}.
\end{equation}
Note that solution for the superconducting gap, $\tilde \Delta_x
(\tilde y)$ of Eq.(36) is not of an exponential shape and changes
its sign several times in space, in contrast to the 3D case
[29,35]. Using Eqs. (34) and (38), we have
\begin{equation}
H^{x}_{\parallel}(0) = 0.815 \ |dH^{x(GL)}_{\parallel}/dT|_{T=T_c}
\ T_c.
\end{equation}
Now, let us summarize the investigation of Eq.(24), which is done
in the same way as our study of Eq.(26). In particular, it is
possible to show that the GL field near transition temperature is
equal to
\begin{equation}
H^{y(GL)}_{\parallel}(T) = \biggl[ \frac{8 \sqrt{2} \pi^2 c
T^2_c}{7 \zeta(3)\sqrt{3} e v_F t_{\perp} c^*} \biggl] \tau ,
\end{equation}
which is $\sqrt{3}$ times smaller than in Eq.(34). Solution of the
Eq.(27) at $T=0$ also gives value of the upper critical field,
which is approximately two times less than the upper critical
field (38), obtained for Eq.(36):
\begin{equation}
H^{y}_{\parallel}(0)= 0.5 \times H^{x}_{\parallel}(0).
\end{equation}
On the above mentioned grounds, we can conclude that the solutions
(21) of Eq.(23) are essential for our calculations of the parallel
upper critical magnetic field in the chiral triplet in-plane
isotropic Q2D superconductor. Therefore, experimentally measured
ratio has to be close to the value (2),(39). It is important that
the theory developed by us above cannot be obtained in the
framework of the Gor'kov [27] and Werthamer-Helfand-Hohenberg [33]
approaches. There are two reasons for that. First, electrons in a
Q2D superconductor in a parallel magnetic field move along open
trajectories in ${\bf p}$ space, in contrast to closed electron
orbits of Refs.[29,35]. Second, we have two-component order
parameter, instead of one-component [29,35].

As we already mentioned, in the candidate for the chiral triplet
in-plane isotropic superconductivity, Sr$_2$RuO$_4$, the
corresponding experimental coefficients [17-19] are almost two
times smaller than the calculated in this Letter (2), which is a
strong argument against the chiral triplet scenario. In this
context, it is important that the parallel upper critical magnetic
field in the above mentioned compound is a good measurable
quantity, unlike the upper critical magnetic fields in some other
Q2D superconductors. In our opinion, the experimental value of the
$H_{\parallel}(0)$ at low temperatures are restricted by the
paramagnetic spin-splitting effects, which are always present in
the singlet and some triplet superconducting phases. This point of
view is also supported by the first order nature of the phase
transition in parallel magnetic fields at low temperatures
[17-19]. The recent Knight shift experiments [15] also strongly
demonstrate the existence of the paramagnetic spin-splitting
effects in superconducting phase of the Sr$_2$RuO$_4$. In
conclusion, let us discuss the physical model we have used for the
calculations of the parallel upper critical magnetic field. It is
known [36] that in the Sr$_2$RuO$_4$ perpendicular to the
conducting layers coherence length (32) is much larger than the
inter-plane distance. Therefore, we have used Eq.(36), instead of
the so-called Lawrence-Doniach model [37,38]. We have also make
use of the fact that metallic phase of the Sr$_2$RuO$_4$ is a good
Fermi liquid [2,3].


The author is thankful to N.N. Bagmet (Lebed) for useful
discussions.

$^*$Also at: L.D. Landau Institute for Theoretical Physics, RAS, 2
Kosygina Street, Moscow 117334, Russia.

\end{document}